\documentclass[preprint,12pt,authoryear]{elsarticle}
\usepackage{amssymb}
\usepackage{siunitx}
\usepackage{textcomp,mathcomp}
\usepackage{graphicx}
\usepackage[hyphens]{url}
\usepackage{hyperref}
\usepackage{cleveref}
\hypersetup{breaklinks=true}

\journal{}

\begin{document}

\begin{frontmatter}

\title{Multi-faceted Impacts of the COVID-19 Outbreak}

\author[label1]{Aneesh Mathews Paul} \author[label2]{Sinnu Susan Thomas}
\address[label1]{Department of Network Planning, BSNL, Ernakulam, Kerala, India, 682020.}
\address[label2]{Image and Vision Computing Lab, Digital University Kerala India.}

\begin{abstract}
COVID-19 has affected our day-to-day life extensively and we present the holistic view on the effects of this pandemic in certain aspects of life. We see a diverse educational system, environmental problems and many religions in India and how they react to this pandemic. We present a survey that revolves over effects on education, environment, and religion mainly concentrated around India.
\end{abstract}
\iffalse
%%Graphical abstract
\begin{graphicalabstract}
\begin{figure}[h]
\centering
\includegraphics[width=0.5\textwidth]{fueldecline.png}
 \caption{Usage of Fuel during Lockdown.} 
 \end{figure} 
 \begin{figure}[h]
\centering
\includegraphics[width=0.5\textwidth]{pmlevel.png}
 \caption{$PM_{2.5}$ Concentration at Different Cities of India.} 
\label{pmlevel}
\end{figure} 
\end{graphicalabstract}

%%Research highlights
\begin{highlights}
\item The pandemic has affected the entire education system and a new era of distance learning has emerged. A review on various education systems during the pandemic is looked out.
\item COVID-19 has benefits in certain areas such as the environment. The environmental effects are discussed. 
\item Overall change in religious practices has changed and we review these aspects in this paper. 
\end{highlights}
\fi
\end{frontmatter}

\section{Introduction}
\label{sec:introduction}
November 2019 has created a history in the records of the World Health Organization (WHO) while introducing COVID-19 disease to the world. The coronavirus was first found in the 1930s in chickens, but then it was found in humans in the 1960s for the first time. Since then, couple of times humans were a victim of this virus such as  SARS-CoV in 2003, HCoV NL63 in 2004, HCoV HKU1 in 2005, MERS-CoV in 2012, and SARS-CoV-2 in 2019 \cite{rockx2020comparative}. COVID-19 belongs to the category of Severe Acute Respiratory Syndrome Coronavirus 2 (SARS-CoV-2) corona virus category.  SARS-CoV-2 is a positive-sense single-stranded RNA virus causing respiratory problems. Humankind has not faced such a disastrous and global health challenge since the Second World War. Wuhan, Hebei province, China became the epicentre to introduce this respiratory disease to the world. 

WHO Director-General Dr Tedros Adhanom Ghebreyesus declared the SARS-CoV-2 outbreak a Public Health Emergency of International Concern on $30^{th}$ January 2020. COVID-19 has changed the entire lifestyle of mankind \cite{haleem2020effects}. Every facet of life has been affected despite being from rich or poor countries \cite{whoCOVID19}. There are positive as well as negative impacts of this epidemic. The world has changed a lot during this time.
A lot of researchers are working on COVID-19, while being a novel problem with many unexplored areas. Researchers have worked on prediction \cite{zheng2020predicting}, classification \cite{wang2020aweakly,oh2020deep}, diagnosis \cite{han2020accurate,ouyang2020dual,kang2020diagnosis}, hardware \cite{ding2020wearable,tripathy2020easyband}. Very little research is being carried out showing the impacts of COVID-19 \cite{chamola2020acomprehensive}. The authors in \cite{chamola2020acomprehensive} has studied the impact of COVID-19 on global economy and explored the use of Internet of Things,  Unmanned Aerial Vehicles, Blockchain, Artificial Intelligence, and 5G to mitigate the impact of COVID-19 outbreak. Devaux et al. \cite{devaux2020new} investigated the effects of hydroxychloroquine against SARS-CoV-2 virus. Faridi \cite{faridi2018middle} has studied the effect of Middle East respiratory syndrome coronavirus (MERS-CoV) that has caused havoc in Saudi Arabia in 2012. The author has seen the effect of MERS-CoV on male and female in Riyadh. The authors in \cite{temsah2020the, kisely2020occurrence,dong2020thesocial,fernandez2020mental} assessed the psychological stress of COVID-19 on health workers. Xiang et al. \cite{xiang2020theimpact} reported an overview of infected healthcare workers in China and Italy during the early periods of the COVID-19. The authors in \cite{ali2020covid, razai2020mitigating, hwang2020loneliness} studied some social impacts of COVID-19. Chakraborty and Maity \cite{chakraborty2020covid} studied the COVID-19 effect on the economy and global environment. Ivanov \cite{ivanov2020predicting} predicted the impact of COVID-19 on global supply chains. Xu et al. \cite{xu2020possible} studied the air quality index to see the effects of COVID-19 on the environment. Chinazzi et al. \cite{Chinazzi2020theeffect} studied the effect of travel and quarantine influence on the dynamics of the spread of COVID-19. Braun \cite{braun2020themoment} narrated examples of the situations of the poor during COVID-19. Ahmed et al. \cite{ahmed2020theprecarious} highlighted the precarious position of postdoctoral fellows in academic positions due to COVID-19. Staniscuaski et al. \cite{Staniscuaski2020impact} projected out the problems faced by academic mothers having many difficulties working at home during COVID-19. Bouillon et al. \cite{bouillon2020coronavirus} discussed the positive side effect of coronavirus on air pollution. Suicide rate has increased during the pandemic time \cite{devitt2020can,hughes2020uncomfortably}.The situation of COVID-19 has diverse effects in India \cite{mukherjee2020covid19}. 

In this paper, we study the multi-faceted effects of COVID-19 on our planet. Our contribution in this paper is threefold.
\begin{enumerate}
\item The pandemic has affected the entire education system and a new era of distance learning has emerged. A review on various education systems during the pandemic is looked out.
\item COVID-19 has benefits in certain areas such as the environment. The environmental effects are discussed. 
\item Overall change in religious practices has changed and we review these aspects in this paper. 
\end{enumerate}

The remainder of this paper is organized as follows. Section 1 highlights the overall change in the education system during the COVID-19 season. Section 2 and 3 presents the environmental and religious effects
of coronavirus respectively. Section 4 presents the conclusions of this paper.

\section{Educational Aspects}
Education system is one of the key pillars in developing a nation. It constitutes an important ingredient in determining the growth of a country. Human development is an important determinant in a person's health and trade. The education system is severely interrupted in most of the countries since the outbreak of this pandemic across the globe. The schools, colleges, and universities are in the total closure mode. Billions of academic learners became devoid of their knowledge acquisition during this pandemic. The teachers, students, schools, and families \textemdash all became a victim of this bitter truth. The world has gone under complete reorganization during this period be it any sector, the education sector is not left apart. The speed of the pandemic and the closure of schools was so fast that it was difficult to come up with a solution with all facilities. The closure of educational institutes will not only have short term impact, but leave a footprint on economic and societal components. There are number of areas in education that is affected by the pandemic:
\begin{enumerate}
\item Cross border movements
 
The landscape of higher education across the world is defined by the cross border movements of the students. Globally every year there is an increase of $10\%$ in the number of students studying abroad as shown in Fig. \ref{noofinternationalstudents}. As per the UNESCO \cite{UNESCO2019}, the students enrolled for higher education for a period of typically a year to seven years. According to the statistics given by Organization for Economic Co-operation and Development (OECD), the international student population with demographic changes is likely to reach 8 million by 2025 \cite{studyabroad}. Most of the international students prefer either the United States, the United Kingdom, Germany, France, or Australia for their higher education \cite{migrationdata} as shown in Fig. \ref{countrystudying}. As per the statistics in 2019, the top host countries involved in sending students to other countries include China, India, South Korea, and France \cite{UNESCO2019} as shown in Fig. \ref{hostcountry}. The pandemic has brought a sluggish impact on the movement of students across the border. The travel restrictions during lockdown and the fear of pandemic will affect the cash flow at the universities. Parents are afraid to send their ward across any border in this situation. The universities in these countries are undergoing extreme pressure on student admission. If this problem persists, there is a possibility of decline in international higher education in the coming years. The pandemic has brought a devastating effect on the global education system. The pandemic has shrinked the world under their own home and hometown and cross border movements seem to be a threat to the life of an individual.
\begin{figure}[h]
\centering
\includegraphics[trim={0 0 0 0},clip,width=0.5\textwidth]{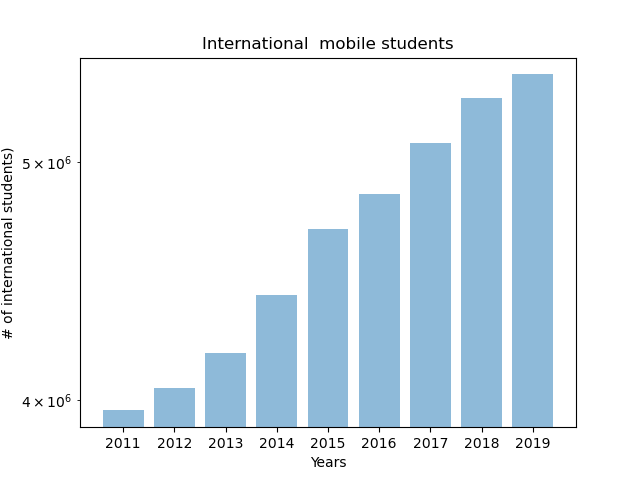}
 \caption{Number of international mobile students.} 
\label{noofinternationalstudents}
\end{figure}

\begin{figure}[h]
\centering
\includegraphics[trim={0 0 0 0},clip,width=0.5\textwidth]{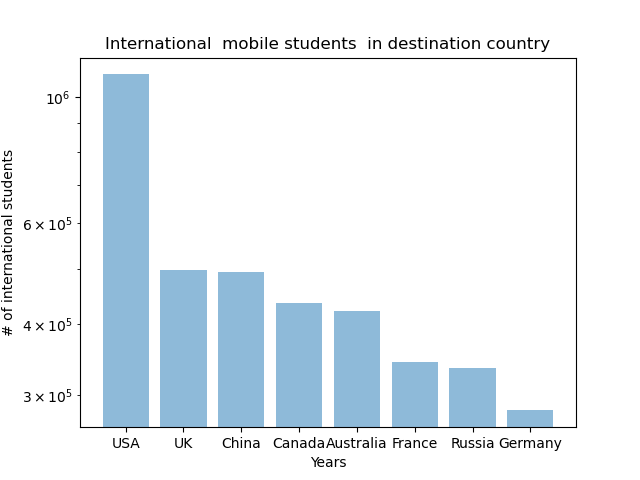}
\caption{The destination country preferred by international mobile students,2019.} 
\label{countrystudying}
\end{figure}
 
\begin{figure}[h]
\centering
\includegraphics[trim={0 0 0 0},clip,width=0.5\textwidth]{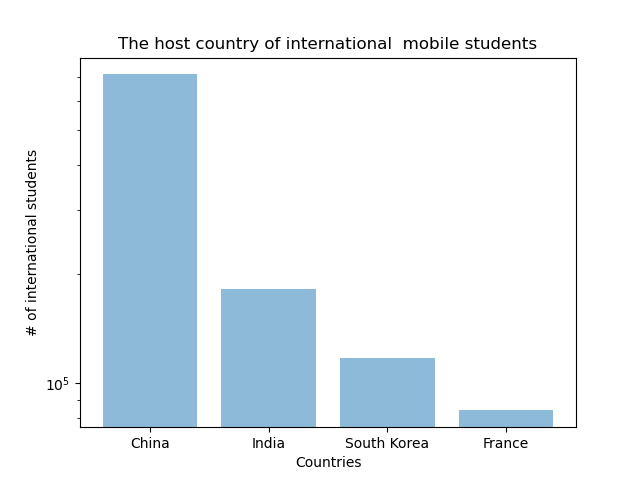}
\caption{The host country of international mobile students, 2019.} 
\label{hostcountry}
\end{figure}
 
\item Online learning
 
Active learning is not only a source of fun but also a source of formation of cognitive social skills. Carlsson et al. \cite{carlsson2015effect} emphasized on the increase in cognitive skills with the total number of school days attended. The study carried out in Sweden showed that crystallized intelligence can be augmented significantly by $1\%$ of a standard deviation while attending ten days of extra schooling. The closure of schools for almost a month at the beginning of this pandemic can cause a trivial loss of $3\%$ of the standard deviation. The pandemic has left the learning systems with no options other than embracing a distant or online learning.

As per the statistics released by UNESCO \cite{UNESCO2019disruption}, the pandemic has affected nearly 1.3 billion learners around the world. The recovery of the disruption of the learning process is essential to facilitate the continuity of the education system. When physical presence is a risky situation, an alternative has to be taken at various levels of learning. Online learning is a new strategy embraced by the education system in this time of pandemic. The transition from active learning to passive learning was very rapid during this pandemic. The curriculum was not designed for passive learning, so the viewers are losing their interest in the content.

The shift in developed countries to the online learning system does not pose any problem, but for developing and under-developed countries - it is a challenging situation. The rural areas of these countries do not have the basic infrastructure to facilitate the online learning. The pandemic has posed a threat to the overall development of the underprivileged in these countries resulting in shattering their economies. 
The video telephony softwares is being used for distant learning. The concept of keeping the electronic gadgets far from the children has been loosened even for a primary school going child during this pandemic. The online learning has removed the commuting time for the learners but on the other hand, made them addicted to electronics devices leading to many social, psychological, and physical disabilities.\\

Online learning brought a paradigm shift in one's own comfort zone. The hassle of traffic jams, pollution, queues, health problems, allergies is halted in this course of time. Most of the learners are happy with the online learning system since environmental problems do not leave them void of attending classes.

\item Unprepared teachers

Online learning has brought an end to the centuries old practice of chalk and talk. Due to the sudden change to the online learning in the education system, the preparedness of the tutors was a concern. An inhibition of this sudden change was found in the tutors during the beginning phase. The course curriculum was not made for passive learning. The sudden shift in the teaching system with inadequate preparation from the learner side was also noticeable. Teaching is a knack that everybody is not gifted with, so many tutors are not so effective in an online mode. In countries like India, where there is a huge shortage of technology savvy tutors, this model of learning would not work out. Lack of infrastructure and resources in the rural parts of these countries is an obstacle for teachers for a complete preparation of imparting the knowledge.

An unavailability of dedicated online platforms is posing a threat for outcome based education. The tutors are adjusting the platforms with the video telephony platforms. If the problem of pandemic persists, there is a need for creating dedicated learning platforms.

\item New admission procedure

Most of the schools and universities undergo the admission process during the month of May-June for Fall Semester. Due to the severity of the pandemic in many parts of the world, the admission process is hindered. The situation in the admission process is becoming alarming in the foreseeable future with the pandemic situation.

Traditional admission procedures would not take place in this season. New procedural strategies for admissions should be considered in order to fill the gap in this pandemic. Some universities are not considering taking any students the current academic year, while some are luring people with discounts. It is a crucial task for the students to decide which school they would like to attend without visiting respective campuses.  

\item Collaborative work

The pandemic has forced people to create a virtual world of working at home. The virtual world creates effortless paths to collaborate across the globe. The conferences, academic meetings, classes, and seminars have gone online leaving a space for academic collaborations. We see a lot of unprecedented collaborative work globally among the educators \cite{honigsfeld2020teacher} during this pandemic leading to a loss in the travel economy. The  cancellation of university-funded international travel for conferences, blanket bans on any international travel for spring break, canceling study-abroad programs \cite{iwai2020} made different academicians closer virtually. Collaborations serve a larger purpose as an individual and also as an organization \cite{green2015interprofessional}. There are lots of scope for online conferencing platform business. The concept of education will be reformed envisaging the global collaboration. Globally, the collaboration has brought a new direction to certification courses
and degrees. These collaborations fulfills the need of each other while dividing the work in chunks.  

\item Recruitment procedure

The pandemic has brought a halt to the organizational structure making a scarcity in the manpower. The universities are facing challenges to recruit new students, and faculty during this pandemic. The retention is also questionable. The recruitment for the faculty is a worrisome issue for the administration when the risk of losing students is hovering around them. When survival of many institutes is a burden for them, the recruitment of new faculty members increases their load. 

Due to the recession in the corporate sector, the recruitment process for the students is a great disaster. The job offers have been withdrawn creating a havoc in the student community. The global outlook of the pandemic would massively devastate the livelihoods in the entire world. Fig. \ref{G7} shows the number of unemployed people in G7 countries according to March 2020. According to the statistics released by Centre for Managing Indian Economy \cite{unemploymentdata} on $1^{st}$ July, 2020, the unemployment rate in India has increased by three times during the pandemic outbreak with urban employment rate shooting to 25.79$\%$ and rural to 22.48$\%$. Fig. \ref{unemployment} shows how the unemployment rate of developing countries have devastated during the pandemic.

\begin{figure}[h]
\centering
\includegraphics[trim={0 0 0 0},clip,width=0.5\textwidth]{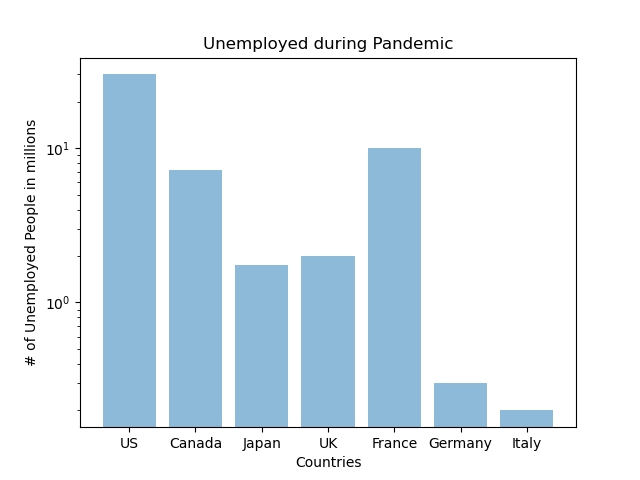}
\caption{No. of Unemployed People in G7 Countries.} 
\label{G7}
\end{figure}

\begin{figure}[h]
\centering
\includegraphics[trim={0 0 0 0},clip,width=0.5\textwidth]{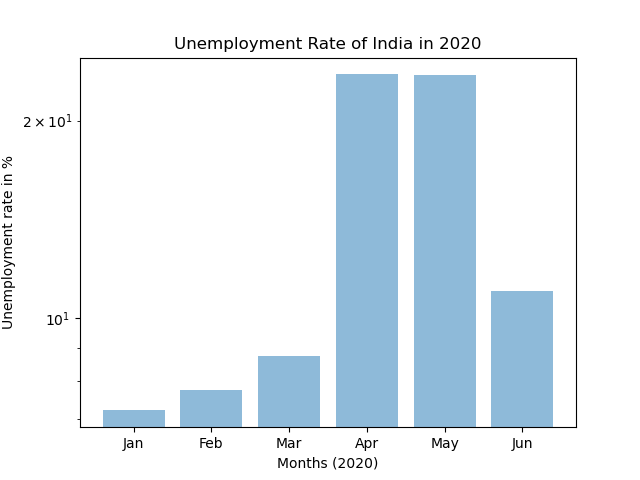}
\caption{Unemployment Rate of India in 2020.} 
\label{unemployment}
\end{figure}

\item Economic impact

The consequences of a sudden shift in the learning system brought a slowdown to the world economy. The international students from China and India constitute 33.7$\%$ and 18.4$\%$ of the total international students in the USA higher education sector. The travel restrictions during the pandemic would cut down the admission process leading to an economic burst-down. The conveyance to the institutes are at a halt causing recession in the travel sector. 

All the learners cannot afford to stay near their institutes, so they stay far and face a time-consuming and costly commute \cite{commutingdata}. Students spend approximately \pounds 100 a month for commutation to their academic institutes. Pandemic has saved the pocket of students in higher education. In countries like India, private schools and private vehicles charge a heavy conveyance fare for the commutation. The pandemic has given relief to the parents. Same time, the train services and the road services are hit badly. Cash-flow in these services reduced leading to an economic crunch in these sectors.

\item Health Benefits

The students use to take long commutes to the institutes taking away their well-being \cite{longcommutingdata}. They are deprived of their sleep and exercises. To commute long distances, students get up early and the daily routine is hampered. Lack of daily exercises make them obese which is a major cause of concern among the youngsters.

Students carry a heavy load of bags on their backs to the school in countries like India. Carrying school bags are back breaking work to the students. Heavy loads of school bags have deleterious effects on the spine of children \cite{malhotra1965carrying}. Many measures are taken to reduce the amount of school baggage, but it was all at a minuscule level. The online learning during the pandemic season turned out to be a heavy relief to the students carrying heavy school bags.

Being in a well-being state is an important aspect of human being. We tend to give rest to the body if it is not in a position to commute. The learners refrain from going to class if they are not well. The pandemic situation takes off all the health issues and helps in smooth learning of classes. The students are free to learn from their home in any physical condition.

\item Assessment procedures

The structure of the learning system is based on various assessment procedures. The students are assessed based on the merit system. The pandemic hit the world during the key assessment period cancelling many exams. The cancellation of exams would have a long-term consequence on the career of the student. First, the internal assessment and then the public examinations were cancelled. The grades at the end of the academic year were predicted according to some undefined rules influencing the privileged students. Education system is shifting to an online assessment system that can create measurement errors. These errors in an abrupt assessment would increase the differences between the privileged and under-privileged students in the future. The labour market would face the dire consequences of inefficient assessment scores. 

The entrance exam in higher education is a worst hit in assessment procedures. The entrance exams to top universities are either postponed or cancelled. The exam agencies are coming up with alternative solutions in consultation with the international institutes.

\item Strike free education

Education system is at stake be it teachers strike or any other political strike. These strikes prevent students from attending classes. According to the study at Argentina \cite{jaume2019long}, 180 days of teachers strike there is a decline of years of education by $3.1\%$ in an academic year. The teacher strike has a negative effect on student learning and their overall achievement \cite{wills2014effects}. Frequent political strikes or hartals impact the overall education system. According to the statistics in Kerala, India \cite{keralahartal2019}, there would be one hartal in every four days leading to disruption in the holistic coverage of prescribed syllabus. 

The online education system is not affected by any sort of socio-political disruptions. Education system in the virtual platform eased out the disturbances due to the strike.

\item Pressure on Family 

The education aspects during the pandemic impacted the family in many ways. 
\begin{enumerate}
    \item The education system comes with mid day meals for the underprivileged in countries like India and the pandemic situation has taken the bread out of the mouth of some children. Children from poor families would come to school with the greed for getting a one time meal. If the pandemic persists, then there is a high chance of drop-outs from the school. Moreover, it would be a tremendous challenge to keep up the motivation of the underprivileged children after the pandemic.
    \item Most of the parents in the pandemic era are working from home. It is difficult for most of the parents to handle domestic pressure and work pressure at home. Working parents are juggling with children and working at home. Global home schooling would produce disparities depending on the ability of the family members to help their children learn. The inequality in each student skill set would overall affect human capital growth.
    \item The unprecedented learning system needs assistance of basic infrastructure for its smooth conduct. Power supply and internet connectivity are the essentials needed without disruption. To avail these resources at home and keep the student without stress is a burdensome work for the parents. In developing countries, it is a difficult situation to maintain the resources around the clock.
    \item Women take care of the children and relatives at home when compared to men. They are more insecure in their jobs.  Women are struggling with their household obligations and work during the pandemic. The juggling between children at home and work would reduce their opportunities and earnings at the workplace. Women have to work harder in order to compensate for the workload and at an increased stress during this period. The study says that many women have left their job during the pandemic due to the imbalance in the worklife. COVID19 is a disaster that would widen the gender inequalities. 
   \end{enumerate} 

\item Loss of societal skills
 
Studies reveal that there is an increase of $60\%$ in usage of electronic gadgets by the impressionable minds. Gadget addiction is one of the major drawbacks of the online learning system. Irritational behavioural patterns are observed in the students during this pandemic. The long time exposure to electronic gadgets are making them obese. An attachment towards gadgets creates a space for emotional imbalances in their personality. 

Students have confined themselves into their own territory keeping them away from the societal component of life. Studying and living together with their companions under one roof increases their social abilities but lockdown has created a void space for problem solving and decision making skills. Social unawareness and lack of cognitive skills would be more visible. These skills improve their employability, productivity, health, and well-being in the future, and ensure the overall progress of the nation.
\end{enumerate}

\section{Environmental Aspects}
\begin{enumerate}
\item Greenhouse gases emission

People around the world are worried about the undergoing changes in the climate. The global temperature is a major concern for many environmental changes. The last five years (2015-2019) were recorded as the hottest years. Globally $3$ \textcelsius $\;$ temperature has increased since the last century. An increase in per capita Gross Domestic Product (GDP) is proportional to global warming. A study conducted by \cite{cederborg2016is} shows the environmental degradation and CO$_2$ emission has increased with the economic growth and more production \cite{book}. According to the census in 2019, the countries with the highest $CO_2$ emission in the world is shown in Fig. \ref{CO2emission}. We see that the environmental degradation increases with the increase in production for economic growth. A lot of measures were taken to reduce the hazardous emission, but a substantial decrement was not possible. The $CO_2$ or greenhouse gas disturbs the natural regulation of temperature in the atmosphere and leads to global warming and climate change.

\begin{figure}[h]
\centering
\includegraphics[trim={0 0 0 0},clip,width=0.5\textwidth]{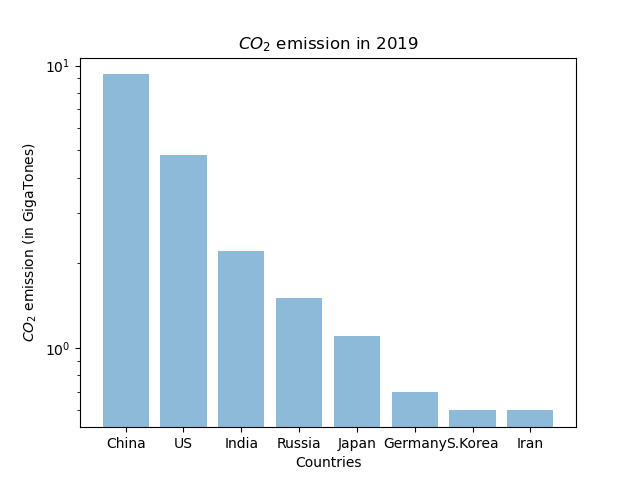}
 \caption{$CO_2$ emission in 2019.} 
\label{CO2emission}
\end{figure}

\begin{figure}[h]
\centering
\includegraphics[trim={0 0 0 0},clip,width=0.5\textwidth]{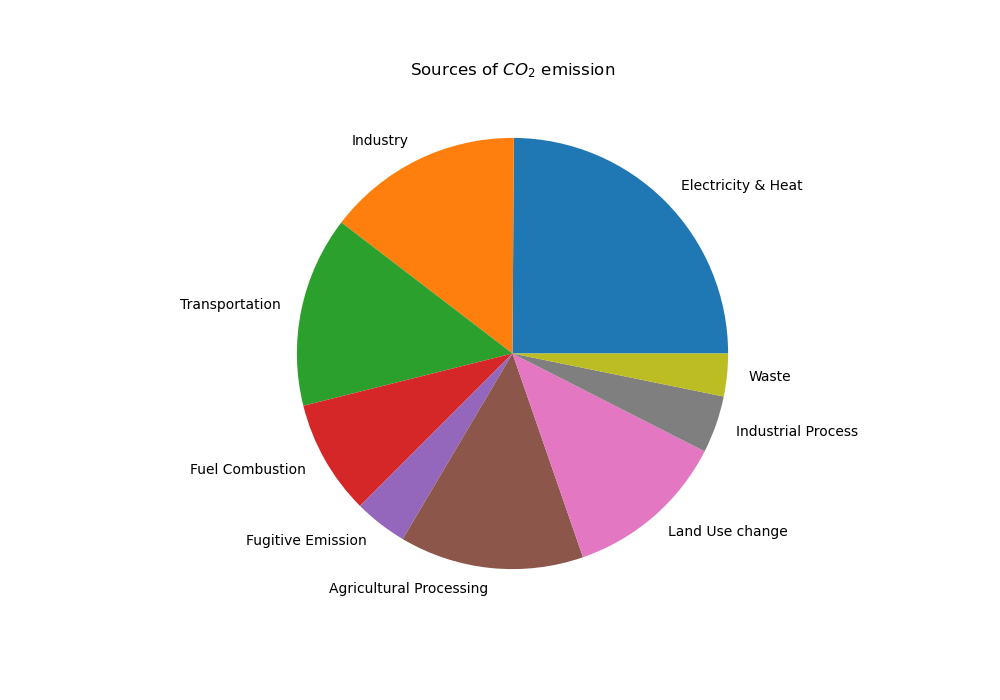}
 \caption{Sources of $CO_2$ emission.} 
\label{sourceofCO2}
\end{figure}

Humans manipulated nature according to his whims and fancies that resulted in paradoxical imbalances. Humans are responsible for the emission of the greenhouse gas in the atmosphere over the last 200 years. COVID19 is the only disaster that has come as a boom to the environment. The major sources of CO$_2$ emission are energy, agricultural processing, land use changes, industrial processing, and other waste. Electricity and heat is generated by burning fossil fuel, coal, and natural gas. A total of 24.9$\%$ greenhouse gases are emitted while burning these fuels and are the leading cause for temperature regulation. Industries emit 14.7$\%$, transportation - 14.3$\%$, agriculture processing - 13.8$\%$, land use change - 12.2$\%$, and industrial processes- 4.3$\%$. Distribution of different sources of greenhouse gas are shown in Fig. \ref{sourceofCO2}.

Before the arrival of pandemic, it was difficult to control the industrial and transportation emission. An impossible action of putting a halt on these hazardous sources was done overnight. According to estimates published by International Energy Agency \cite{iea2020global} in mid April, global $CO_2$ emission are reduced by 8$\%$ or 2.6 GigaTonnes (GT) which is equivalent to a decade earlier data. There was an average decline of 25$\%$ energy demand per week during full lockdown and an average of 18$\%$ decline in partial lockdown countries. An unprecedented decline in demand for various fuels is seen during the pandemic as shown in Fig. \ref{fueldecline}. The crisis of pandemic is paving a way for clean energy transitions. This decline in $CO_2$ emission is unprecedented and would be temporary, unless there is a resilient effort to change the structure.  

\begin{figure}[h]
\centering
\includegraphics[trim={0 0 0 0},clip,width=0.5\textwidth]{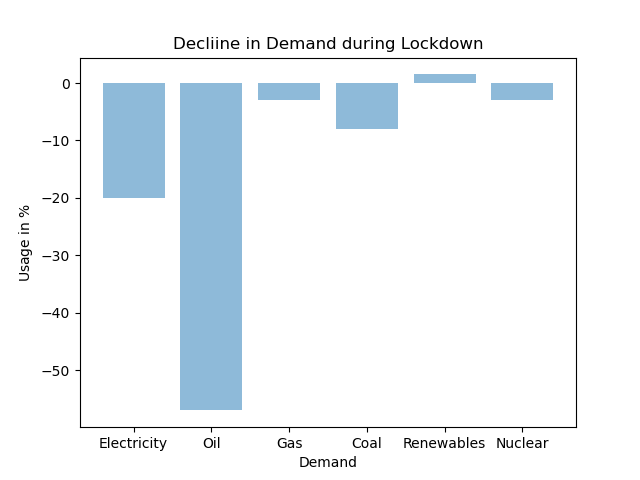}
 \caption{Usage of Fuel during Lockdown.} 
\label{fueldecline}
\end{figure}

\item Vibration in the earth crust

High frequency seismic waves are propagated into the earth mainly due to the activities of the human. The seismic noise renders the real time estimate of population dynamics. The COVID19 pandemic period is the longest seismic noise quiet period ever recorded. According to the Royal Observatory of Belgium \cite{Lecocqeabd2438}, the seismic noise of the earth during the pandemic is not prevalent, reducing the vibration of the earth by $50\%$. The vibrations are reduced by one-third of the normal activity during the lockdown. It becomes easy for seismologists to detect the movement in the earth crust without much of an expedition.

\item Construction Projects

The construction projects in some countries were at complete hold during the initial stages of the lockdown. The availability of the workforce and the site constraints halted some of the projects. Construction activities create an adverse impact on the environment. The burning of fossil fuel, noise,  and the waste of the construction  contribute to the regulation of the temperature in the environment. The halt in construction reduced the amount of PM10 by three times in the month of April 2020. Air pollution is recorded highest in many cities of India. The annual average PM2.5  concentration during the lockdown was much better than the safer limit \cite{delhidata,mumbaidata,bengalurudata,chennaidata} as shown in Fig. \ref{pmlevel}.

\begin{figure}[h]
\centering
\includegraphics[trim={0 0 0 0},clip,width=0.5\textwidth]{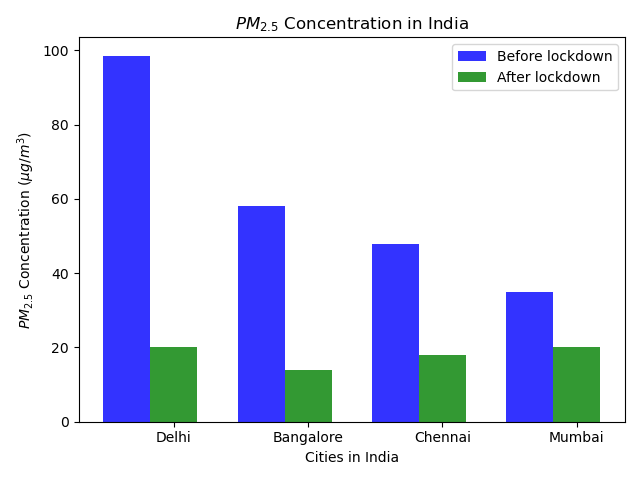}
 \caption{$PM_{2.5}$ Concentration at Different Cities of India.} 
\label{pmlevel}
\end{figure} 

\item Breathing rivers

Under the banner of economic growth, entire industrial and other waste is dumped into the rivers making it difficult to breathe. The aquatic species are becoming extinct due to the pollutants in the river.  India is at the top of river pollution. Ganges river is the most populated river in the world. The present pandemic has come as a blessing in disguise for rivers. The water pollution has decreased considerably during COVID19 period. The waters from the rivers in India are tested during COVID19 and the results provoke us to take measures to clean the rivers. The pH levels, the conductivity level, dissolved oxygen (DO), and the biological oxygen demand (BOD) of the water is reduced during the lockdown period \cite{arif2020reduction}. A betterment in standards of drinking water was seen during the lockdown period as shown in Fig. \ref{drinkingwaterstandard}.

\begin{figure}[h]
\centering
\includegraphics[trim={0 0 0 0},clip,width=0.5\textwidth]{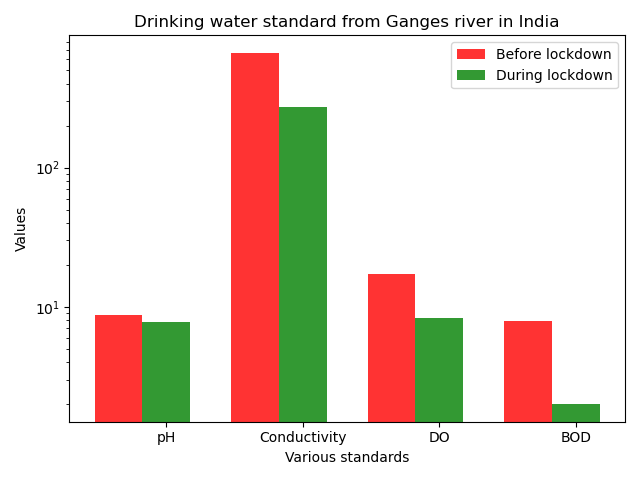}
 \caption{Drinking Water Standards of River Ganges in India.} 
\label{drinkingwaterstandard}
\end{figure}

\item Wildlife effects

Pandemic season was a lockdown for mankind, but on the contrary animals were liberalized. Humans were away but animals took over the deserted cities and towns. Animals took the advantage of the drop in human activity and came out to explore and play in the public places. Scavengers are not around to shoo them away giving a space for wildlife to thrive. Mallard ducks, wild deers, herd of goats, troop of monkeys, kangaroos, gangs of turkeys, and many others are taking human spaces. Road mortality was a threat to the wildlife population \cite{ucd2020impact}. The mortality has reduced to $58\%$ due to less traffic on roads in the USA. Less roadkill reduces the ecological imbalances. 

Some animals have successfully adapted to live alongside humans and their survival is dependent on them. An absence of human activity endangers some wildlife species. Some governments mobilize funds to feed and preserve these animals, and the lockdown hindered their progress. According to the Livestock Census of 2012, there are around 18 million stray dogs in India. These dogs are fed by NGOs or leftovers from restaurants. The closed restaurants and the restrictions in the movement made these stray dogs starve.

The sustenance of the people in rural places of poor countries became difficult during the pandemic. People are driven to take extreme steps for their livelihood through poaching. The illegal hunting of endangered species in African continent is a threat for the wildlife society. According to study conducted by TRAFFIC, the wildlife poaching in India has increased twice during the pandemic period. It has increased from $22\%$ to $44\%$ during the lockdown period. It may turn disastrous and pave a way for another pandemic. Humans struggled from recent pandemics such as AIDS, EBOLA, MERS, and SARS that came as an effect of consumption of animal meat \cite{salyer2017prioritizing,machalaba2015anthropogenic}. It becomes the responsibility of the Wildlife Conservation Society to prevent any pandemic in the future.

\item Non-COVID diseases

Due to the clean air and lockdown, non-COVID diseases are at steep decline in countries prone to all pollution. The behavioural changes during the lockdown has brought a decline in insurance claims by $40\%$ in India. Waterborne infectious diseases and respiratory related diseases are being recorded as lowest during the pandemic time. The claims on deadly diseases such as cancer has turned down by $42\%$ as per the statistics of the insurance companies in India \cite{medicalclaim2020}.

Due to decrease in vaccination \cite{WHO2020} and disruption in the hospital services, there is a possibility of an outbreak of other diseases.
\end{enumerate}

\section{Religious Aspects}

Religion makes people follow different practices and form socio-cultural groups. Each culture recorded in human history practised some organized system of beliefs and practices. We tend to see very few people practicing faith in normal life. For some it seems absolutely mandatory but for some these are obnoxious practices. Religion and faith is an integral part of people's lives worldwide, even though it is increasing. Religious practices were hampered during lockdown. Various aspects of religion during lockdown are discussed in detail: 
\begin{enumerate}
    \item Religious Polarization
    
    Religion is a predominant factor for satisfaction in life, on the contrary the religious tensions can be annoying \cite{migheli2019religious} and affect the economic growth of the country. Religious fervency is vigorous in most secularized countries \cite{ribberink2018religious}. The polarization towards targeted groups increased in many countries during the earlier stage of the pandemic \cite{niranjan2020india}. Since the cases of the virus were aggravated by the religious gathering in some countries, we could see religious bigotries coupled with the pandemic. The virus has morphed itself into an anti- community virus \cite{ipsos2020,canada2020}. The bigotries and xenophobia towards different sects of people can be seen in different countries as shown in Fig. \ref{xenophobia}.
    \begin{figure}[h]
    \centering
    \includegraphics[trim={0 0 0 0},clip,width=0.5\textwidth]{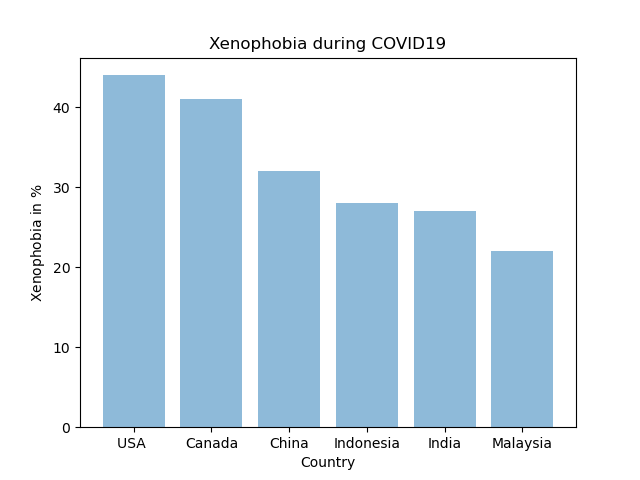}
     \caption{Percentage of xenophobia during COVID-19.} 
    \label{xenophobia}
    \end{figure}
    
    Religion and politics are a crucial part of life and COVID-19 has acquainted the human life without these jargon words. The places that culminated religious polarization at the earlier stages of the pandemic were felt at peace in the later stages of the pandemic. Everybody came out in unison to curb this pandemic through their services. Charity works and social commitment was seen at large during the pandemic.
    
    \item Life without Religious Practices
    
    The role of religious practices in spreading COVID-19 was predominant \cite{quadri2020covid,wesley2020religion}. The religious leaders surpassing the mass gathering orders became a source of virus carriers in the entire nation. Some of the early COVID19 outbreaks were traced back to religious gatherings such as Daegu church in South Korea, Bnei-Brak in Israel, Oom in Iran, Tablighi-Jamaat in India, Tabligh-e-Jamaat in Malaysia, prayer meeting in France, and many more \cite{dave2020god,schuchat2020public,james2020high}. The pandemic spread in various countries was sparked by religious gatherings as shown in Fig. \ref{religiousspread}. In South Korea $63.5\%$, Israel $40\%$, USA $38\%$, Malaysia $36\%$, India $26\%$, and Pakistan $22\%$ COVID19 cases can be traced back to religious gatherings.
    \begin{figure}[h]
    \centering
    \includegraphics[trim={0 0 0 0},clip,width=0.5\textwidth]{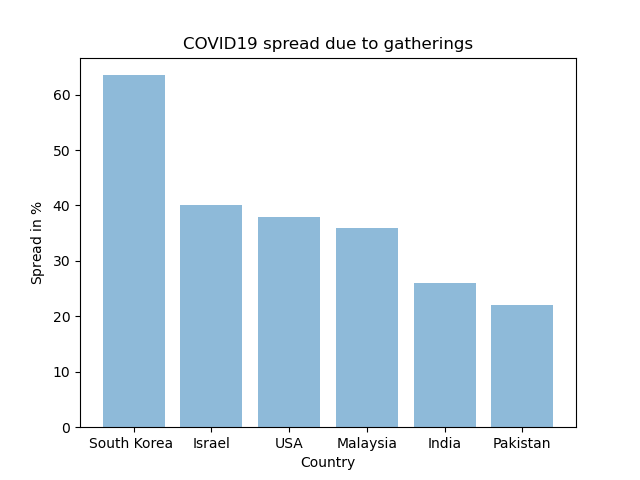}
     \caption{COVID19 Spread due to Religious Gatherings.} 
    \label{religiousspread}
    \end{figure}
    
    Many people are fervent in religious practices such as visiting places of worship, mass gatherings, religious celebrations, and many more. All these practices are hindered during the pandemic. Entire paradigm shift was seen in the religious fraternity. The religious holidays and celebrations were practiced at home. The key moments of rituals were experienced in their own home. Religious leaders were bound to ask their followers to stay at home during pandemics. They started releasing double the amount of messages  for the community to cope up with the stress during the pandemic. The religious organizations started doing more charitable services. People started living with faith rather than religious places. Social distancing would be the most tricky in places of worship. The survey concludes that the public has become comfortable staying at home and practising their faith till the resumption of the normal situation \cite{prerna2020have}.     
    
    \item Technical Advancements
    
    Religious leaders are challenged to foster and to bring their services and communities together in these trying times from a distance. The online platforms were used to connect to the community during religious ceremonies. During the pandemic time, the searches for prayer have skyrocketed in Google search engines. Many spiritual and therapeutic activities, such as yoga, meditation, martial arts, and conscious dance classes have gone online during this pandemic. These temporary solutions are not sustainable solutions as they need physical relationships with people. 
    
    The places of worship is a source of income for many religious leaders and the common man. These sources of income are hindered by the pandemic. Life without religious practices also hit livelihoods of businesses around the places of worship.
    \begin{figure}[h]
    \centering
    \includegraphics[trim={0 0 0 0},clip,width=0.5\textwidth]{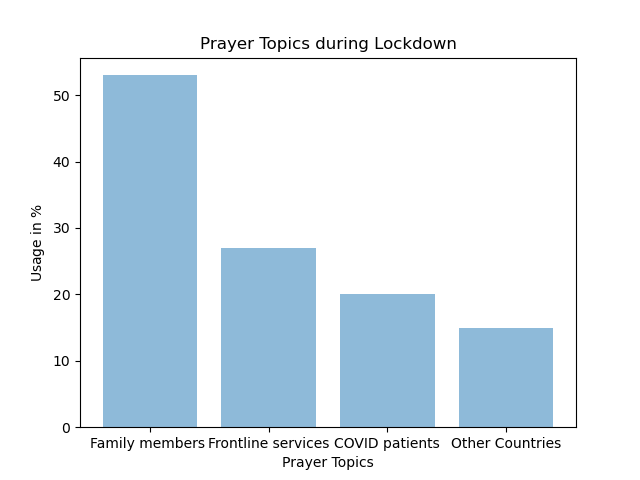}
     \caption{Prayer Topics During Lockdown.} 
    \label{prayertopics}
    \end{figure}
    \item Path towards Prayer
    
    A loneliness during the pandemic times created furore among the individuals. People were complacent in their comfort zone but they were kicked out of that with hopelessness and despair. Adapting to a new environment with a U-turn in an individual's life was a difficult task. Life is fragile during pandemic time but increase in spirituality and faith became a vital part of their life. Religion is considered as a source of solace in terms of pain and scepticism. The role of prayer in the current pandemic situation among the general public is noteworthy \cite{bentzen2020crisis}. There was an increased interest ever recorded in search of prayer as per the daily data recorded from Google for 95 countries. According to Tearfund COVID19 Prayer Public Omnibus Research \cite{tearfund2020} conducted in the UK during the lockdown period gauged the responses to spiritual practices. The statistics was conducted on 2,101 UK adults aged 18+  and shows that nearly half ($44\%$) of UK adults pray regularly and a quarter ($24\%$) of UK adults attended online religious service during lockdown. One in twenty UK adults ($5\%$) who attended religious service have never gone to church and Two-thirds ($66\%$) of UK adults agree that prayer changes the world. Generally, religion is more appealing to the older generation, but during the lockdown period the religious revival was seen in younger ones. The highest number of Quran apps from Google Playstore was downloaded during pandemic \cite{how2020}. Irrespective of any religion, everybody started seeking hope in their faith and started praying for various topics as shown in Fig. \ref{prayertopics}.
    
\end{enumerate}

\section{Conclusion}
We humans have gone through multiple virus pandemics in different times. Pandemic came with human devastation but with times we came over it. COVID-19 is a disaster in many aspects of life, but in some it has proved a blessing. This paper describes the multiple faces of virus outbreak. We have looked upon a few possible areas of life which have been affected by COVID-19 such as the educational sector, environmental sector, and religious sector. The areas where it is a boom leaves a space to ponder on the living standard of human beings. Lot of effort was taken with respect to some serious problems on the earth, but everything was in vain and it was noticed that there was a sudden break in these problems during a pandemic. Once the pandemic is over, there is a call by the earth to make it a better healthy living place.

\bibliographystyle{elsarticle-harv}
\bibliography{references}

\end{document}